# Theoretical investigation on the transition metal borides with Ta$_3$B$_4$-type structure: a class of hard and refractory materials

Naihua Miao, Baisheng Sa, Jian Zhou, Zhimei Sun,*

*Department of Materials Science and Engineering, College of Materials, Xiamen University, 361005 Xiamen, People's Republic of China*

*Author to whom correspondence should be addressed. Tel and Fax: +86-592-2186664. Email addresses: zmsun@xmu.edu.cn and zhmsun2@yahoo.com.

**Abstract**

Based on density functional theory, we have systematically studied the structural stability, mechanical properties and chemical bonding of the transition metal borides M$_3$B$_4$ (M=Ti, V, Cr, Zr, Nb, Mo, Hf, Ta, and W) for the first time. All the present studied M$_3$B$_4$ have been demonstrated to be thermodynamically and mechanically stable. The bulk modulus, shear modulus, Young's modulus, Poisson's ratio, microhardness, Debye temperature and anisotropy have been derived for ideal polycrystalline M$_3$B$_4$ aggregates. In addition, the relationship between Debye temperature and microhardness has been discussed for these isostructral M$_3$B$_4$. Furthermore, the results of the Cauchy pressure, the ratio of bulk modulus to shear modulus, and Poisson's ratio suggest that the valence electrons of transition metals play an important role in the ductility of M$_3$B$_4$. The calculated total density of states for M$_3$B$_4$ indicates that all these borides display a metallic conductivity. By analyzing the electron localization function, we show that the improvement of the ductility in these M$_3$B$_4$ might attribute to the decrease of their angular bonding character.

## 1. Introduction

Transition metal borides (TMBs) have been the focus of attention during the recent years due to their useful mechanical and electrical properties and their potential industrial applications (e.g. cutting tools and coatings). Among these TMBs, $ReB_2$ and $OsB_2$ are ultra-incompressible and superhard [1-3]; $MgB_2$ has been observed to be a superconductor at 39 K [4]; $HfB_2$, $ZrB_2$, MoB, $CrB_2$, CrB, $TiB_2$, $Ti_3B_4$ and TiB have been used as coatings or reinforcements in various composites because of their great hardness and good thermal stability [5-13]. As far as we know, most of the experimental works and theoretical studies are concentrated on the transition metal monoborides and diborides. While, another class of TMBs ($M_3B_4$) with $Ta_3B_4$-type structure (Space Group IMMM, No.71) also exhibit great hardness [14-16], possess high melting point[16-18], show good electrical conductivity [14, 17, 18], and display considerable oxidation-resistant in air [16-18], which can be regarded as potentially useful materials for high temperature engineering applications, but insufficient attention was paid to them. Theoretical works on the electronic and bonding properties of $Ta_3B_4$-type compounds have been carried out, but they were mainly performed for $Ta_3B_4$ [17]. A computational thermodynamic study on the stability of $V_3B_4$ suggested that it was stable enough as a refractory material for high temperature composite applications [19]. Experimentally, it has been shown that the platelet morphology of the $Ti_3B_4$ phase could improve the strength and toughness of the platelet composite [20], and the $Ti_3B_4$ and $TiB_2$ peritectic composite particulates have been in-situ synthesized to reinforce metal matrix composite [11]. Large crystals of $Cr_3B_4$, $Ta_3B_4$ and related solid solutions have been prepared and measured to have a Vickers microhardness ($H$v) around 22 GPa by Shigeru Okada *et al.* [14, 16]. A relatively larger *H*v value of 33±2 GPa has also been reported for $Ta_3B_4$, indicating its potential applications for nanostructured superhard materials [15]. However, up to now, due to the limited published works on $M_3B_4$, the knowledge of these materials is rather scarce, e.g., their structural stability, mechanical properties and chemical bonding are still unknown. Therefore, a systematic investigation on this class of hard and refractory materials is of great practical interest and importance.

Computational technique based on density functional theory is a powerful tool to explore the phase stability and physical and mechanical properties of materials. In the present work, by means of density functional theory calculations, we focused on transition metal borides $M_3B_4$ (M=Ti, V, Cr, Zr, Nb, Mo, Hf, Ta, and W). Among these borides, $Hf_3B_4$ and $W_3B_4$ have not yet been reported experimentally and their structural stabilities are still unknown. Hence, we began with our study on the thermodynamic stabilities of $M_3B_4$ (Section 3.1), and then systematically explored their mechanical properties (Section 3.2), Debye temperature (Section 3.3), anisotropy (Section 3.4), electronic structure and chemical bonding (Section 3.5), and dynamical stability (Section 3.6). Our results will provide a fundamental understanding on these hard and refractory materials and offer reference data for experimentalists. The remainder of the paper is organized as follows. In Section 2, we describe the computational methods and details. In Section 3, the results and discussions are presented. Finally, the conclusions are given in Section 4.

## 2. Calculation methods

The present calculations are based on the density functional theory (DFT). And there are 14 atoms for each cell, i.e., 8 boron atoms and 6 transition metal atoms.

Firstly, the Cambridge Serial Total Energy Package (CASTEP) [21] was used for geometry optimizations and elastic constants calculations. The interactions between the ions and the electrons were described by using the Ultrasoft Vanderbilt pseudopotentials (USPP) [22]. The valence electrons were treated as: $2s2p$ for B; $3s3p3d4s$ for Ti, V and Cr; $4s4p4d5s$ for Zr, Nb and Mo; $5d6s$ for Hf, Ta; and $5s5p5d6s$ for W. The generalized gradient approximation (GGA) [23] and Perdew Burke Ernzerhof functional (PBE) [24] were employed for exchange correlation functional. The structures were optimized through the Broyden–Fletcher–Goldfarb–Shanno (BFGS) method [25]. A cutoff energy of 600 eV was chosen for plane wave basis set and the 12×4×12 k-points were automatically generated by Monkhorst-Pack (MP) scheme [26]. The convergence criteria were set to maximum force tolerance within 0.03 eV/Å for geometry optimization, and 0.006 eV/Å for elastic constants calculations.

To obtain the force constants for phonon calculations and to check the reliability of results gained by CASTEP, the Vienna *ab initio* simulation package (VASP) code as implemented to solve the Kohn-Sham equations employing the projector augmented wave (PAW) method [27-30] was also used. The semi-core $p$ states and possibly the semi-core $s$ states were treated as valence states. PAW-GGA-PBE [23, 24, 29, 30] pseudopotentials with electronic configurations of B $2s^22p^1$, Ti $3p^63d^24s^2$, V $3p^63d^34s^2$, Cr $3p^63d^54s$, Zr $4s^24p^65s^24d^2$, Nb $4p^65s^24d^3$, Mo $4p^65s^14d^5$, Hf $5p^66s^25d^2$, Ta $5p^66s^25d^3$, and W $5p^66s^25d^4$ were employed. The cutoff energy for plane wave basis set was 700 eV. 12×4×12 k-points and MP scheme [26] were adopted for Brillouin Zone sampling. The relaxation convergence for ions and electrons were $1\times10^{-5}$ eV and $1\times10^{-6}$ eV, respectively. The force constants, lattice parameters, density of states, and electron localization function (ELF) [31, 32] analyzed by VESTA [33] were calculated for the equilibrium structures.

Our phonon calculations were performed through the supercell approach [34]. Force constants of supercells were prepared by using the VASP, and the PHONOPY code [35, 36] was used to calculate the phonon frequencies and phonon density of states.

## 3. Results and discussions

### 3.1. Lattice parameters and thermodynamic stability

The calculated lattice parameters of $M_3B_4$ are given in Table. 1. Note that the lattice constants predicted by CASTEP and VASP are in good agreement with each other and the available experimental data, indicating the reliability of the results obtained from the present density functional theory approaches. As seen from Table. 1, the predicted lattice constants of $Ti_3B_4$, $V_3B_4$ and $Nb_3B_4$ are in good agreement with the experiments [37-39]. For $Ta_3B_4$, the calculated lattice parameters are a bit larger than the experimental data (e.g., about 1.6% larger for *b*) [16], while for $Cr_3B_4$, the calculated lattice constants are slightly smaller than the experiments (e.g., about 2.7% smaller for *a*) [40], which is within the GGA error. However, for the other compounds $M_3B_4$ such as $Zr_3B_4$, $Mo_3B_4$, $Hf_3B_4$ and $W_3B_4$, there are no theoretical or experimental data available for comparison. Therefore, the present results could provide useful information for further experimental or theoretical investigations. From Table. 1, it is obvious that the lattice constants (*a*, *b* and *c*) follow the orders of $Ti_3B_4<V_3B_4<Cr_3B_4$ for 3*d* $M_3B_4$, $Zr_3B_4<Nb_3B_4<Mo_3B_4$ for 4*d* $M_3B_4$, and $Hf_3B_4<Ta_3B_4<W_3B_4$ for 5*d* $M_3B_4$, which is reasonable as the radii of elements in the same row of the periodic table decrease with increasing atomic numbers. Moreover, we also observed that the predicted lattice parameters for $Zr_3B_4$ and $Hf_3B_4$ were a bit larger than the other $M_3B_4$.

This can be understood as Zr and Hf possesses larger atomic radii than the other transition metals studied herein.

To investigate the possibility to obtain $M_3B_4$, we have calculated the formation energy $E_{form}$ by Eq. (1) according to the reaction (2).

$$E_{form} = E_{total}^{(M3B4)} - 3E_{total}^{(M)} - 4E_{total}^{(B)} \quad (1)$$

$$3M + 4B \rightarrow M_3B_4 \quad (2)$$

The total energy $E_{total}$ of M and B were calculated for stable crystalline transition metals and α-boron, respectively. The results are also given in Table. 1. For all the $M_3B_4$ borides, the formation energies are negative, suggesting that all of them are thermodynamically stable. Hence, we expect that all of them, including two new compounds $Hf_3B_4$ and $W_3B_4$, could be realized experimentally.

### 3.2. Mechanical properties

The mechanical properties of solids are of great importance because they relate to many properties of materials, e.g. interatomic potentials, equation of states, melting points and phonon spectra. Here, we start with the calculations of elastic constants.

### 3.2.1. Elastic constants

For orthorhombic $M_3B_4$ crystals, there are nine independent elastic constants which are usually referred to as $c_{11}$, $c_{22}$, $c_{33}$, $c_{44}$, $c_{55}$, $c_{66}$, $c_{12}$, $c_{13}$ and $c_{23}$. In the present work, we obtained the elastic constants by applying a given homogeneous deformation (strain) and calculating the resulting stress [41] as implemented in CASTEP. The calculated elastic constants are presented in Table. 2. Among these borides, $V_3B_4$ exhibits the largest elastic constants of $c_{22}$, $c_{33}$, $c_{44}$, $c_{55}$, and $c_{66}$, while $W_3B_4$ possesses the largest elastic constants of $c_{11}$, $c_{12}$, $c_{13}$ and $c_{23}$. As seen from Table. 2, $c_{11}$, $c_{12}$, $c_{13}$ and $c_{23}$ elastic constants increase with increasing valence electrons concentration (VEC), which indicates the valence electrons of transition metals in $M_3B_4$ may play an important role in these elastic constants. For orthorhombic crystals, the Born mechanical stability criteria are given by: $c_{11}>0$, $c_{22}>0$, $c_{33}>0$, $c_{44}>0$, $c_{55}>0$, $c_{66}>0$, $(c_{11}+c_{22}-2c_{12})>0$, $(c_{11}+c_{33}-2c_{13})>0$, $(c_{22}+c_{33}-2c_{23})>0$, $(c_{11}+c_{22}+c_{33}+2c_{12}+2c_{13}+2c_{23})>0$. It is obvious that all of the studied $M_3B_4$ crystals with $Ta_3B_4$-type structure satisfy the Born criteria, hence they are all mechanically stable.

### 3.2.2. Mechanical parameters for polycrystalline aggregates

From the calculated elastic constants, other mechanical parameters for polycrystalline aggregate such as bulk modulus ($K$), shear modulus ($G$), Young's modulus ($E$), and Poisson's ratio ($\sigma$) can be derived using Voigt–Reuss–Hill (VRH) approximation [42-44]. Voigt's and Reuss's schemes represent the upper and lower bounds to the elastic modulus, respectively. For orthorhombic crystals, the shear modulus ($G$) and the bulk modulus ($K$) according to Voigt and Reuss approximations are defined as:

$$G_V = \frac{1}{15}(c_{11}+c_{22}+c_{33}-c_{12}-c_{13}-c_{23}) + \frac{1}{5}(c_{44}+c_{55}+c_{66}) \quad (3)$$

$$K_V = \frac{1}{9}(c_{11}+c_{22}+c_{33}) + \frac{2}{9}(c_{12}+c_{13}+c_{23}) \quad (4)$$

$$G_R = \frac{15}{4(s_{11}+s_{22}+s_{33})-4(s_{12}+s_{13}+s_{23})+3(s_{44}+s_{55}+s_{66})} \quad (5)$$

$$K_R = \frac{1}{(s_{11}+s_{22}+s_{33})+2(s_{12}+s_{13}+s_{23})} \quad (6)$$

The Hill's averages are taken from the averages of the two [44]:

$G = (G_V + G_R)/2 \quad (7)$

$K = (K_V + K_R)/2 \quad (8)$

Then the Young's modulus ($E$), and Poisson's ratio ($\sigma$) are calculated by the following formulas:

$E = 9KG/(3K + G)$     (9)

$\sigma = (3K - 2G)/2(3K + G)$     (10)

Based on equations (3)-(10), we derived the corresponding values which are presented in Table. 3. As far as we know, there are no experimental results available related to the elastic modulus for $M_3B_4$ studied here, but we can compare these modulus with other known ceramic crystals. As seen from Table. 3, the bulk modulus $K$ follow the orders of $Ti_3B_4 < V_3B_4 < Cr_3B_4$ for $3d$ $M_3B_4$, $Zr_3B_4 < Nb_3B_4 < Mo_3B_4$ for $4d$ $M_3B_4$, and $Hf_3B_4 < Ta_3B_4 < W_3B_4$ for $5d$ $M_3B_4$, indicating that the increased extra valence electrons of transition metals contribute to the chemical bonding of $M_3B_4$. And for all studied borides, their shear modulus ranging from 165 GPa to 238 GPa may reveal that there are strong directional bonding in these borides. Among them, $V_3B_4$ shows the largest shear modulus (238 GPa) and Young's modulus (555 GPa). The shear modulus of $V_3B_4$ is much larger than that of α-$Al_2O_3$ (143 GPa) [45] and is comparable to that of stishovite (218 GPa) [45] and γ-$Si_3N_4$ (248 GPa) [45]. A largest bulk modulus value of 335 GPa for $W_3B_4$ is also observed in Table. 3, suggesting that $W_3B_4$ is very uncompressible, which is about 11% lower compared with c-BN (376 GPa) [45] but much higher than $Ti_{0.25}Al_{0.75}N$ (178 GPa, the benchmark cutting tool material today) [46]. It is also noted in Table. 3 that, for all the borides studied here, their Poisson's ratio ($\sigma$) are very small, indicating all of them are relatively stable against shear. As is known, the range of Poisson's ratio for central-force solids are 0.25−0.5 [47, 48], thus, the interatomic forces are central in $Mo_3B_4$ and $W_3B_4$, while they are non-central in the other $M_3B_4$ as their Poisson's ratio are all smaller than 0.25.

### 3.2.3. Microhardness

To estimate the microhardness ($H_V$) of $M_3B_4$, we used the following relation for isotropic solids:

$$H_V = (1-2\sigma)E/[6(1+\sigma)] \qquad (11)$$

The estimated microhardness $H_V$ is also listed in Table. 3. Our estimated $H_V$ values are ranging from 18.3 ($W_3B_4$) to 26.4 ($V_3B_4$) GPa indicating that all these $M_3B_4$ are hard materials. The calculated microhardness 23.9 GPa for $Cr_3B_4$ is in good agreement with the experimental value 21.9±1.0 GPa [18]. For $Ta_3B_4$, Shigeru Okada *et al* suggested that it had a same level of microhardness as $Cr_3B_4$ [16], which is consistent with our prediction. Nevertheless, when compared with the reported $H_V$ 33±2 GPa for $Ta_3B_4$ [15], our calculated $H_V$ is a bit smaller. However, it should be noted that our estimated $H_V$ 22.5 GPa for $Ta_3B_4$ is in the range of the values from 21.9 to 33±2 GPa. From Table. 3, we can also observe that $V_3B_4$ shows the greatest microhardness value among these $M_3B_4$ borides. According to Teter [49], the polycrystalline shear modulus is a better predictor of hardness than bulk modulus, the microhardness can be calculated by: $H_T = 0.1769G-2.899$. The result is also given in Table. 3, which is a bit larger than the data obtained from equation (11). Note that the later estimated microhardness $H_T$ for $Ta_3B_4$ is in good agreement with the reported value 33±2 GPa [15]. For both estimated microhardness $H_V$ and $H_T$ of $M_3B_4$, the trends are quite similar, i.e., the higher shear modulus, the higher microhardness.

**3.2.4. Ductility and brittleness**

To explore the ductility and brittleness of $M_3B_4$, we refer to the Cauchy pressure and the ratio of bulk modulus to shear modulus ($K/G$). Pettifor [50] has suggested that the angular character of atomic bonding in metals and compounds, which also relates to their brittleness or ductility, could be described by the Cauchy pressure. For directional bonding with angular character, the Cauchy pressure is negative, with larger negative pressure representing a more directional character, which will be further explored in Section 3.5. Generally speaking, a positive Cauchy pressure reveals damage tolerance and ductility of a crystal, while a negative one demonstrates brittleness. In Fig. 1, the Cauchy pressures ($c_{23} - c_{44}$) for orthorhombic crystals as a function of valence electrons concentration (VEC) are illustrated for $M_3B_4$. Clearly, for all $M_3B_4$, where transition metal elements M are in the same rows (3$d$, 4$d$, and 5$d$), the Cauchy pressures increase with increasing VEC, i.e., $c_{23} - c_{44}$ follow the orders of $Ti_3B_4 < V_3B_4 < Cr_3B_4$ for 3$d$ $M_3B_4$, $Zr_3B_4 < Nb_3B_4 < Mo_3B_4$ for 4$d$ $M_3B_4$, and $Hf_3B_4 < Ta_3B_4 < W_3B_4$ for 5$d$ $M_3B_4$, indicating that the increased valence electrons of transition metals contribute to the ductility of $M_3B_4$. As seen from Fig. 1, it is obvious that the Cauchy pressure for $W_3B_4$ is positive, suggesting the ductility of $W_3B_4$, while for the other studied borides, their Cauchy pressures are all negative, showing that all of them have a brittle nature.

To qualify whether a material would fail in a ductile or brittle manner, Pugh proposed the ratio of bulk modulus to shear modulus ($K/G$) [51]. The transition from ductile to brittle behavior occurs around a $K/G$ value of 1.75. Based on this assumption, materials with $K/G$ values larger than 1.75 are associated with ductility, whereas materials with values smaller than 1.75 correspond to a brittle nature. The plot of $K/G$ ratios as a function of valence electrons concentration (VEC) for $M_3B_4$ is given in Fig. 2. Compared with Fig. 1, a similar trend can be observed for $K/G$ ratios and VEC, i.e., $K/G$ ratios increase with VEC and also follow the orders of $Ti_3B_4 < V_3B_4 < Cr_3B_4$ for 3$d$ $M_3B_4$, $Zr_3B_4 < Nb_3B_4 < Mo_3B_4$ for 4$d$ $M_3B_4$, and $Hf_3B_4 < Ta_3B_4 < W_3B_4$ for 5$d$ $M_3B_4$. The calculated $K/G$ value for $W_3B_4$ is 1.97, obviously larger than 1.75, indicating the ductile nature of $W_3B_4$. This is in good agreement with the above discussion on the ductility of $W_3B_4$. For the other $M_3B_4$, their $K/G$ values are all more or less smaller than 1.75, hence they can be classified as brittle materials.

As is known, Poisson's ratio has great correlation with ductility of crystalline alloys and amorphous metals, and it has been used as a screening parameter to identify intrinsic ductility of metals and alloys [52], i.e., the higher Poisson's ratio, the better ductility at low temperature, and vice versa. The Poisson's ratio as a function of valence electrons concentration (VEC) for $M_3B_4$ has also been given in Fig. 3. The trend for the relationship between Poisson's ratio and VEC is quite similar to that in Fig. 1 and Fig. 2, i.e., Poisson's ratio of $M_3B_4$ goes up with the increase of their VEC. As seen from Fig. 3, $Ti_3B_4$ possesses the smallest Poisson's ratio, and hence it shows the greatest brittleness among these $M_3B_4$ compounds. While $W_3B_4$ exhibits the largest Poisson's ratio indicating its greatest ductility. Furthermore, the brittleness of $M_3B_4$ follow the orders of $Ti_3B_4>V_3B_4>Cr_3B_4$ for 3$d$ $M_3B_4$, $Zr_3B_4>Nb_3B_4>Mo_3B_4$ for 4$d$ $M_3B_4$, and $Hf_3B_4>Ta_3B_4>W_3B_4$ for 5$d$ $M_3B_4$. These results coincide well with the above analysis of Cauchy pressures and $K/G$ values.

### 3.3. Debye temperature

### 3.3.1. The calculation of Debye temperature

Mechanical properties can be related to thermodynamical parameters such as Debye temperature, specific heat, thermal expansion and melting point [48]. We have calculated the Debye temperature ($\theta_D$) for $M_3B_4$ by the relation between the mean sound velocity $v_m$ and $\theta_D$ [53]:

$$\theta_D = \frac{h}{k}\left[\frac{3n}{4\pi}\left(\frac{N_A \rho}{M}\right)\right]^{1/3} v_m \qquad (12)$$

where $h$ is Planck's constant, $k$ is Boltzmann's constant, $N_A$ is Avogadro's number, $\rho$ is the density, $M$ is the molecular weight and $n$ is the number of atoms in the molecule.

The mean sound velocity of polycrystalline materials can be obtained according to the following approximations [53]:

$$v_m = \left[\frac{1}{3}\left(\frac{2}{v_t^3} + \frac{1}{v_l^3}\right)\right]^{-1/3} \quad (13)$$

where $v_l$ and $v_t$ are the longitudinal and transverse elastic wave velocity of the polycrystalline material which can be obtained by using the polycrystalline shear modulus $G$ and the bulk modulus $K$ from Navier's equation [54]:

$$v_l = \left(\frac{K + \frac{4G}{3}}{\rho}\right)^{1/2} \quad (14)$$

$$v_t = \left(\frac{G}{\rho}\right)^{1/2} \quad (15)$$

The calculated results are listed in Table. 4. Among these materials, $V_3B_4$ displays the highest Debye temperature of 1058 K, and about half of this value is obtained for $W_3B_4$. For $Nb_3B_4$, a theoretical $\theta_D$ of 592 K has been reported by Blinder and Bolgar [55], which is 209 K smaller than our predicted value 801K. Nevertheless, it should be noted that, by using the same approach [55], they also obtained other $\theta_D$ values of 339 K for $HfB_2$ (580 K [56]), 440 K for $NbB_2$ (720 K [56]) and 332 K for $TaB_2$ (570 K [56]), which are much smaller than the experimental data (where the experimental values are in the parenthesis). As their theoretical approach generally underestimated the Debye temperature, it is understandable that our calculated $\theta_D$ are higher than their predicted values. Therefore, although there is no other theoretical or experimental data available for comparison, we believe that our predicted $\theta_D$ for $M_3B_4$ are reasonable and can be used as reference for the experimentalists or theorists. For materials, usually, the higher Debye temperature, the larger microhardness.

Interestingly, in the present work, $Hf_3B_4$, $Ta_3B_4$ and $W_3B_4$ show lower Debye temperature compared with $Zr_3B_4$, but all of them are harder than $Zr_3B_4$, which is quite different from the normal cases. Thus, it is worthy to discuss the relationship between $\theta_D$ and $H_V$ which will be given in the following subsection.

### 3.3.2. The relationship between Debye temperature and microhardness

The strength of interatomic cohesive forces in solids is exhibited by such properties as compressibility, microhardness, and melting point [48, 57]. The stronger these forces are, the higher the Debye temperature is; and vice versa [58]. For isostructural groups of crystals, the relationship between Debye temperature ($\theta_D$) and microhardness ($H_V$) can be defined as [59]:

$$\theta_D = p \times H_V^{1/2} V^{1/6} M^{-1/2} + q \quad (16)$$

where $M$ is molar mass, $V$ is molecular volume, and $p$ and $q$ are linear fitted parameters for specified crystal system. This relationship for $M_3B_4$ borides is illustrated in Fig. 4, where the data are taken from the preceding subsections. As seen from Fig. 4, the above relationship has been perfectly represented by using our predicted Debye temperature $\theta_D$ and microhardness $H_V$. The fitted parameters are 1826.13 for $p$, and 7.40 for $q$. And then these parameters were used to recalculate the $H_V$ values for $M_3B_4$ by using the equation (16). Little deviation has been observed in Fig. 4 when these newly obtained microhardness values were compared with the $H_V$ values in Table. 3. So according to the above relation (16), one can roughly estimate $H_V'$ ($\theta_D'$) for a class of isostructural crystals by some given data, i.e., we can firstly obtain the parameters ($p$, $q$) by fitting several pairs of $\theta_D$ and $H_V$, and then calculate corresponding $H_V'$ ($\theta_D'$) with a known $\theta_D'$ ($H_V'$) value.

### 3.4. Anisotropy

The anisotropy of crystals affects the physical properties in different directions of solids, e.g. microcracks are induced in ceramics due to their anisotropy of the coefficient of thermal expansion and elastic anisotropy. Therefore, it is necessary to investigate the elastic anisotropy of $M_3B_4$ borides. The shear anisotropic factors, which provide a measure of the degree of anisotropy in the bonding between atoms in different planes, are given by:

$$A_1 = \frac{4c_{44}}{c_{11} + c_{33} - 2c_{13}} \quad \text{for the \{100\} plane} \quad (17)$$

$$A_2 = \frac{4c_{55}}{c_{22} + c_{33} - 2c_{23}} \quad \text{for the \{010\} plane} \quad (18)$$

$$A_3 = \frac{4c_{66}}{c_{11} + c_{22} - 2c_{12}} \quad \text{for the \{001\} plane} \quad (19)$$

The calculated values of $A_1$, $A_2$ and $A_3$ for $M_3B_4$ are listed in Table. 5. For an isotropic crystal, the values of $A_1$, $A_2$ and $A_3$ equal to 1, while any value smaller or greater than 1 is a measure of the degree of shear anisotropy possessed by the crystal. As seen from Table. 5, none of the shear anisotropic factors comes up to 1, indicating the low anisotropy of $M_3B_4$ borides. Among them, $W_3B_4$ exhibits the largest anisotropic for all three planes, i.e., $W_3B_4$ possesses the highest degree of shear anisotropy for its {100}, {010} and {001} planes. Similar to $W_3B_4$, $Mo_3B_4$ also exhibits large anisotropic factors for all three planes. On the contrary, $Ti_3B_4$, $V_3B_4$ and $Cr_3B_4$ display low anisotropy for all their three planes. $Zr_3B_4$, $Nb_3B_4$, $Mo_3B_4$, $Hf_3B_4$, $Ta_3B_4$ and $W_3B_4$ show larger anisotropy in the {001} plane than their {100} and {010} planes.

In additions, we have also adopted another way to measure the elastic anisotropy by using the percentage of anisotropy in the compression and shear introduced by Chung and Buessum [60], which are defined as:

$$A_K = \frac{K_V - K_R}{K_V + K_R} \qquad (20)$$

$$A_G = \frac{G_V - G_R}{G_V + G_R} \qquad (21)$$

For crystals, a value of zero denotes elastic isotropy and a value of 100% represents largest anisotropy. The percentage of bulk and shear anisotropies are also given in Table. 5. We observed that all the $M_3B_4$ possessed low bulk anisotropy as the values ranging from 0 to 0.5%. It is obvious that $W_3B_4$ and $Mo_3B_4$ exhibit relatively high shear anisotropies among these borides, while the values of their bulk anisotropies are small.

**3.5. Electronic structure and chemical bonding**

In order to gain better understanding of the electronic structure and chemical bonding of $M_3B_4$, we have calculated the total density of states (TDOS), which are presented in Fig. 5. The calculated TDOS of $Ta_3B_4$ agrees well with the previous work [17]. Note that the TDOS of $M_3B_4$ in the same column show similar characters, and there are finite values at the Fermi Levels, indicating the metallic conductivity of $M_3B_4$ as observed by experiments [14, 18]. For all studied borides, their Fermi Levels locate at the shoulders of the peaks of TDOS, suggesting that all of them are stable [61], which is consistent with the previous discussion on their stability. Moreover, the Fermi Levels of $M_3B_4$ move from a lower energy level to a higher one, following the orders of $Ti_3B_4<V_3B_4<Cr_3B_4$, $Zr_3B_4<Nb_3B_4<Mo_3B_4$, and $Hf_3B_4<Ta_3B_4<W_3B_4$, which indicates that the increased valence electrons of transitions metals contribute to the bonding states in $M_3B_4$ and hence enhance their bulk modulus.

To further explore the chemical bonding of $M_3B_4$, the topological analysis of the electron localization function (ELF) [31, 32] has been carried out, since it gives a rather quantitative picture on the chemical bonding of compounds and provides a convenient mathematical framework enabling an unambiguous characterization of bonds. Contour plots of ELF on the (100) plane of $M_3B_4$ are illustrated in Fig. 6. Clearly, the $B_1$-$Hf_1$, $B_1$-$B_2$ and $B_2$-$B_3$ covalent bonding of $Hf_3B_4$ with angular character can be observed in Fig. 6 (a), where the maximum ELF values between the bonded atoms are greater than 0.75. Moreover, as seen in Fig. 6 (a), two labeled regions $P_1$ and $P_2$ with relatively high ELF values are visible, and the ELF values of $Hf_2$-$Hf_3$ and $Hf_2$-$Hf_1$ bonding are obviously smaller than that of $Hf_1$-$Hf_3$ bond, which also indicates angular bonding character around Hf atoms in $Hf_3B_4$. A quite different chemical bonding character for $Ta_3B_4$ and $W_3B_4$ can be seen in Fig. 6 (b) and (c). The angular $B_1$-$Ta_1$, $B_1$-$B_2$ and $B_2$-$B_3$ covalent bonding in $Ta_3B_4$ are weakened compared with that in $Hf_3B_4$, and they are further weakened in $W_3B_4$, as characterized by the maximum ELF values or regions between the bonded atoms decreasing from $Hf_3B_4$ to $Ta_3B_4$ and to $W_3B_4$. Furthermore, when comparing Fig. 6 (b) with Fig. 6 (a), it is observed that the $P_1$ region of $Ta_3B_4$ almost disappears and its $P_2$ region becomes weaker. While in Fig. 6 (c), both regions $P_1$ and $P_2$ vanish and the bonding among the nearest W atoms become homogeneous, suggesting a non-angular character of W-W bonding in $W_3B_4$. Therefore, the angular character of B-B, M-B, M-M bonding is weakened from $Hf_3B_4$ to $Ta_3B_4$ and to $W_3B_4$ with the increasing of the extra valence electrons. The similar trend can be observed from the ELF plots for the other $M_3B_4$, where the angular bonding character of $M_3B_4$ follow the orders of $Ti_3B_4$>$V_3B_4$>$Cr_3B_4$, $Zr_3B_4$>$Nb_3B_4$>$Mo_3B_4$, and $Hf_3B_4$>$Ta_3B_4$>$W_3B_4$. Hence, the decrease of angular bonding character in $M_3B_4$ might be the reason for the increase of Cauchy pressure and the improvement of their ductility as seen in Section 3.2.

### 3.6. Dynamical stability

In the preceding subsections, we have demonstrated that all of the studied $M_3B_4$ are thermodynamically and mechanically stable. To further investigate the dynamical stability of two new predicted compounds $Hf_3B_4$ and $W_3B_4$, we calculated phonon dispersions for them, which are illustrated in Fig.7 (a) and (b). The phonon dispersion curves of $Hf_3B_4$ and $W_3B_4$ are a little complex and show different characters. It is known that negative or imaginary frequencies indicate the dynamical instability of crystals. As seen from the phonon dispersion curves, no negative frequency have been found for $Hf_3B_4$ or $W_3B_4$, suggesting both compounds are dynamically stable.

## 4. Concluding remarks

In summary, by means of density functional theory calculations, we have systematically study the structural stability, mechanical properties, electronic structure and chemical bonding of $M_3B_4$ (M are IVB, VB and VIB transition metals, i.e., M=Ti, V, Cr, Zr, Nb, Mo, Hf, Ta, and W). All the studied transition-metal borides with $Ta_3B_4$-type structure are thermodynamically and mechanically stable. The predicted lattice parameters and microhardness for them are in good agreement with the available experimental results.

The bulk modulus, Young's modulus, shear modulus, Poisson's ratio, microhardness and Debye temperature have been derived from the calculated elastic constants for ideal polycrystalline $M_3B_4$ aggregates. Among them, $V_3B_4$ exhibits the highest shear modulus (238 GPa), Young's modulus (555 GPa), microhardness (26.4 GPa) and Debye temperature (1059 K), while $W_3B_4$ possesses the largest bulk modulus (335 GPa) but the lowest Debye temperature (532 K). In addition, the relationship between Debye temperature and microhardness has been discussed for these isostructral $M_3B_4$. Furthermore, by analyzing the Cauchy pressure, the ratio of bulk modulus to shear modulus, and Poisson's ratio, we found that the ductility of $M_3B_4$ follows the orders of $Ti_3B_4<V_3B_4<Cr_3B_4$, $Zr_3B_4<Nb_3B_4<Mo_3B_4$, and $Hf_3B_4<Ta_3B_4<W_3B_4$, suggesting that the valence electrons of transition metals play an important role in the ductility of $M_3B_4$, i.e., the more valence electrons, the better ductility.

Moreover, the calculated total density of states of $M_3B_4$ indicates that all these borides display a metallic conductivity, and the increased valence electrons of transitions metals contribute to the bonding states in $M_3B_4$ and hence enhance their bulk modulus. By analyzing the electron localization function, in the view of chemical bonding, we show that the decrease of angular bonding character in $M_3B_4$ might be the reason for the increase of Cauchy pressure and the improvement of their ductility. Finally, the calculated phonon dispersions of the new predicted compounds $Hf_3B_4$ and $W_3B_4$ suggest that both of them are dynamically stable. We expect our results provide a fundamental understanding on these hard and refractory materials and offer reference data for further investigation or applications of this class of transition metals borides.

**Acknowledgements**

This work is supported by National Natural Science Foundation of China (60976005), the Outstanding Young Scientists Foundation of Fujian Province of China (2010J06018) and the program for New Century Excellent Talents in University (NCET-08-0474). The State Key Laboratory for Physical Chemistry of Solid Surfaces at Xiamen University is greatly acknowledged for providing the computing resources.

**FIGURE CAPTIONS:**

Fig. 1. (Color online) Plot for Cauchy pressure as a function of valence electron concentration (VEC).

Fig. 2. (Color online) Plot for $K/G$ as a function of valence electron concentration (VEC).

Fig. 3. (Color online) Plot for Poisson's ratio as a function of valence electron concentration (VEC).

Fig. 4. The relationship between Debye temperature and microhardness $H_V$. The data points in squares are calculated using the data from the preceding sections.

Fig. 5. The calculated total density of states for $M_3B_4$. The Fermi Levels have been set to 0 eV and marked by short dash lines.

Fig. 6. (Color online) Contour plots of ELF on the (100) plane of $M_3B_4$ for (a) $Hf_3B_4$, (b) $Ta_3B_4$, (c) $W_3B_4$. The color scale for the ELF value is given at the left of the figure, where all the mappings are under the same saturation levels and the interval between two nearest contour lines is 0.13.

Fig. 7. The calculated phonon dispersion curve for (a) $Hf_3B_4$ and (b) $W_3B_4$.

Table. 1. The calculated lattice parameters (*a*, *b* and *c*) and formation energy ($E_{form}$) for $M_3B_4$. US and PAW denote the results obtained from CASTEP and VASP; Exp. represents the experimental data.

| $M_3B_4$ | | a(Å) | b(Å) | c(Å) | $E_{form}$ (eV/cell) |
|---|---|---|---|---|---|
| $Ti_3B_4$ | US | 3.260 | 13.733 | 3.036 | -13.470 |
| | PAW | 3.263 | 13.755 | 3.042 | - |
| | Exp.[37] | 3.259 | 13.730 | 3.032 | - |
| $V_3B_4$ | US | 3.041 | 13.200 | 2.976 | -11.826 |
| | PAW | 3.043 | 13.221 | 2.981 | - |
| | Exp.[38] | 3.03 | 13.18 | 2.986 | - |
| $Cr_3B_4$ | US | 2.918 | 13.004 | 2.936 | -6.870 |
| | PAW | 2.923 | 13.030 | 2.942 | - |
| | Exp.[40] | 2.99 | 13.010 | 2.949 | - |
| $Zr_3B_4$ | US | 3.539 | 14.903 | 3.206 | -12.071 |
| | PAW | 3.553 | 14.943 | 3.217 | - |
| | Exp. | - | - | - | - |
| $Nb_3B_4$ | US | 3.307 | 14.103 | 3.144 | -11.494 |
| | PAW | 3.325 | 14.179 | 3.159 | - |
| | Exp.[39] | 3.296 | 14.094 | 3.151 | - |
| $Mo_3B_4$ | US | 3.161 | 13.900 | 3.074 | -6.561 |
| | PAW | 3.175 | 13.974 | 3.086 | - |
| | Exp. | - | - | - | - |
| $Hf_3B_4$ | US | 3.522 | 14.731 | 3.205 | -11.382 |
| | PAW | 3.502 | 14.679 | 3.188 | - |
| | Exp. | - | - | - | - |
| $Ta_3B_4$ | US | 3.332 | 14.227 | 3.176 | -10.183 |
| | PAW | 3.309 | 14.091 | 3.145 | - |
| | Exp.[16] | 3.291 | 13.994 | 3.133 | - |
| $W_3B_4$ | US | 3.185 | 13.923 | 3.073 | -4.054 |
| | PAW | 3.196 | 13.993 | 3.083 | - |
| | Exp. | - | - | - | - |

Table. 2. The calculated elastic stiffness constants $c_{ij}$ (in GPa) for $M_3B_4$.

| $M_3B_4$ | $c_{11}$ | $c_{22}$ | $c_{33}$ | $c_{44}$ | $c_{55}$ | $c_{66}$ | $c_{12}$ | $c_{13}$ | $c_{23}$ |
|---|---|---|---|---|---|---|---|---|---|
| $Ti_3B_4$ | 424 | 520 | 581 | 229 | 242 | 214 | 108 | 94 | 49 |
| $V_3B_4$ | 484 | 640 | 631 | 239 | 264 | 236 | 137 | 140 | 96 |
| $Cr_3B_4$ | 492 | 609 | 612 | 231 | 238 | 205 | 169 | 166 | 159 |
| $Zr_3B_4$ | 372 | 357 | 458 | 184 | 162 | 183 | 114 | 79 | 78 |
| $Nb_3B_4$ | 473 | 546 | 544 | 201 | 242 | 236 | 157 | 168 | 126 |
| $Mo_3B_4$ | 484 | 476 | 559 | 208 | 219 | 196 | 227 | 202 | 193 |
| $Hf_3B_4$ | 419 | 432 | 523 | 208 | 225 | 212 | 128 | 117 | 90 |
| $Ta_3B_4$ | 479 | 559 | 552 | 196 | 232 | 233 | 170 | 182 | 150 |
| $W_3B_4$ | 507 | 459 | 556 | 205 | 223 | 200 | 271 | 225 | 253 |

Table. 3. The calculated Young's modulus ($K_V$, $K_R$, and $K$ in GPa), shear modulus ($G_V$, $G_R$, and $G$ in GPa), elastic modulus ($E$ in GPa), Poisson's ratio ($\sigma$) and microhardness ($H_V$ and $H_T$ in GPa) for polycrystalline $M_3B_4$ aggregates.

| $M_3B_4$ | $K_V$ | $K_R$ | $K$ | $G_V$ | $G_R$ | $G$ | $E$ | $\sigma$ | $H_V$ | $H_T$ |
|---|---|---|---|---|---|---|---|---|---|---|
| $Ti_3B_4$ | 225 | 224 | 224 | 222 | 217 | 220 | 497 | 0.131 | 24.4 | 36.0 |
| $V_3B_4$ | 278 | 275 | 277 | 240 | 236 | 238 | 555 | 0.166 | 26.4 | 39.2 |
| $Cr_3B_4$ | 300 | 298 | 299 | 216 | 213 | 215 | 520 | 0.210 | 23.9 | 35.1 |
| $Zr_3B_4$ | 192 | 191 | 192 | 167 | 163 | 165 | 385 | 0.166 | 18.3 | 26.3 |
| $Nb_3B_4$ | 274 | 273 | 274 | 210 | 205 | 208 | 497 | 0.197 | 23.1 | 33.8 |
| $Mo_3B_4$ | 307 | 306 | 307 | 184 | 177 | 181 | 453 | 0.254 | 20.1 | 29.1 |
| $Hf_3B_4$ | 227 | 226 | 227 | 198 | 194 | 196 | 457 | 0.165 | 21.8 | 31.8 |
| $Ta_3B_4$ | 288 | 288 | 288 | 205 | 201 | 203 | 493 | 0.215 | 22.5 | 33.0 |
| $W_3B_4$ | 336 | 335 | 335 | 177 | 163 | 170 | 437 | 0.283 | 18.9 | 27.2 |

Table. 4. The calculated density ($\rho$ in g/cm$^3$), the longitudinal, transverse and mean elastic wave velocity ($v_l$, $v_t$ and $v_m$ in m/s), and the Debye temperature ($\theta_D$ in K) for M$_3$B$_4$.

| M$_3$B$_4$ | $\rho$ | $v_l$ | $v_t$ | $v_m$ | $\theta_D$ |
|---|---|---|---|---|---|
| Ti$_3$B$_4$ | 4.57 | 10642 | 6935 | 7603 | 1055 |
| V$_3$B$_4$ | 5.45 | 10437 | 6607 | 7268 | 1059 |
| Cr$_3$B$_4$ | 5.94 | 9928 | 6012 | 6645 | 991 |
| Zr$_3$B$_4$ | 6.22 | 8133 | 5148 | 5663 | 735 |
| Nb$_3$B$_4$ | 7.29 | 8687 | 5334 | 5887 | 801 |
| Mo$_3$B$_4$ | 8.14 | 8203 | 4713 | 5234 | 732 |
| Hf$_3$B$_4$ | 11.56 | 6499 | 4118 | 4530 | 591 |
| Ta$_3$B$_4$ | 12.93 | 6572 | 3959 | 4378 | 590 |
| W$_3$B$_4$ | 14.50 | 6227 | 3426 | 3819 | 532 |

Table. 5. The calculated shear anisotropic factors ($A_1$, $A_2$, and $A_3$) for different planes and the percentage of anisotropy in the compression and shear ($A_K$ and $A_G$).

| $M_3B_4$ | $A_1$ | $A_2$ | $A_3$ | $A_K$ (%) | $A_G$ (%) |
|---|---|---|---|---|---|
| $Ti_3B_4$ | 1.122 | 0.964 | 1.175 | 0.306 | 0.989 |
| $V_3B_4$ | 1.143 | 0.978 | 1.110 | 0.415 | 0.856 |
| $Cr_3B_4$ | 1.195 | 1.053 | 1.073 | 0.415 | 0.598 |
| $Zr_3B_4$ | 1.095 | 0.982 | 1.462 | 0.191 | 1.107 |
| $Nb_3B_4$ | 1.182 | 1.153 | 1.336 | 0.055 | 1.070 |
| $Mo_3B_4$ | 1.300 | 1.347 | 1.554 | 0.095 | 1.948 |
| $Hf_3B_4$ | 1.175 | 1.164 | 1.425 | 0.222 | 1.169 |
| $Ta_3B_4$ | 1.177 | 1.145 | 1.334 | 0.107 | 0.990 |
| $W_3B_4$ | 1.335 | 1.753 | 1.876 | 0.083 | 4.045 |